\documentclass[svgnames,usenames,preprint,nocopyrightspace]{sigplanconf}

\usepackage{amsmath}
\usepackage{amssymb}

\usepackage{tikz}
\usetikzlibrary{arrows}
\usepackage{xspace}
\usepackage[colorlinks=true]{hyperref}

\usepgflibrary{shapes.geometric}
\usepgflibrary[shapes.geometric]
\usetikzlibrary{shapes.geometric}
\usetikzlibrary[shapes.geometric]
\usetikzlibrary[matrix]
\usetikzlibrary{arrows, decorations, decorations.markings}
\usetikzlibrary{positioning}

\usepackage{listings}

\newcommand*\lstinputpath[1]{\lstset{inputpath=#1}}
\lstinputpath{code}

\usepackage{array,arydshln}

\usepackage{xcolor}

\definecolor{dkgreen}{rgb}{0,0.6,0}
\definecolor{gray}{rgb}{0.5,0.5,0.5}
\definecolor{mauve}{rgb}{0.58,0,0.82}
\definecolor{light-red}{rgb}{1.0,0.8,0.8}
\definecolor{yellow-orange}{rgb}{1.0,0.85,0.05}
\definecolor{gray-blue}{rgb}{0.57,0.72,0.84}
\definecolor{light-gray}{gray}{0.80}

\definecolor{DarkGreen}{rgb}{0.0, 0.2, 0.13}

\definecolor{apricot}{rgb}{0.98, 0.81, 0.69}
\definecolor{bananayellow}{rgb}{1.0, 0.88, 0.21}
\definecolor{babyblueeyes}{rgb}{0.63, 0.79, 0.95}

\usepackage{soul}
\sethlcolor{titlebgcolor}

\usepackage{wrapfig}
\usepackage{times} 
\usepackage{stmaryrd} 
\usepackage{centernot}

\usepackage{comment}
\excludecomment{enclosedreminders}

\usepackage[
     disable,
    textwidth=0.3\textwidth,textsize=scriptsize]{todonotes}

\newcommand{\mclcomm}[1]{\todo[color=Beige,bordercolor=DarkGoldenrod,linecolor=DarkGoldenrod]{MCL: #1}}
\newcommand{\mclcommin}[1]{\todo[inline,color=Beige,bordercolor=DarkGoldenrod,linecolor=DarkGoldenrod]{MCL: #1}}

\newcommand{\jmcommin}[1]{\todo[inline,color=blue!20,bordercolor=blue,linecolor=blue,size=\scriptsize, caption={}]{JM: #1}}

\definecolor{dkgreen}{rgb}{0,0.6,0}
\definecolor{gray}{rgb}{0.5,0.5,0.5}
\definecolor{mauve}{rgb}{0.58,0,0.82}

\lstdefinestyle{cstyle}{
  frame=tb,
  escapechar={@},
  language=c,
  aboveskip=3mm,
  belowskip=3mm,
  showstringspaces=false,
  columns=flexible,
  basicstyle={\footnotesize\ttfamily},
  numbers=none,
  numberstyle=\tiny\color{gray},
  keywordstyle=\color{blue},
  keywordstyle=[2]\color{dkgreen},
  keywordstyle=[3]\color{magenta},
  commentstyle=\color{gray},
  stringstyle=\color{mauve},
  breaklines=true,
  breakatwhitespace=true,
  tabsize=3,
}

\lstdefinestyle{hstyle}{
  frame=tb,
  escapechar={$},    
  mathescape,         
  language=haskell,
  aboveskip=3mm,
  belowskip=3mm,
  showstringspaces=false,
  columns=flexible,
  basicstyle={\scriptsize\ttfamily},
  numbers=none,
  numberstyle=\tiny\color{gray},
  keywordstyle=\color{blue},
  keywordstyle=[2]\color{magenta},
  commentstyle=\color{gray},
  stringstyle=\color{red},
  breaklines=true,
  breakatwhitespace=true,
  tabsize=3,
  morekeywords={},
  keywords=[2]{CAssign, CAddAssOp,CAddOp,CBinary,CAssignOp, CCompound,
    CFor, CLeOp, CBlockStmt, CUnary, CPostIncOp, undefNode, CExpr, CMulOp}
}

\lstdefinestyle{cstyleTable}{
  frame=tb,
  escapechar={@},
  language=c,
  aboveskip=3mm,
  belowskip=3mm,
  showstringspaces=false,
  columns=flexible,
  basicstyle={\small\ttfamily},
  numbers=none,
  numberstyle=\footnotesize\color{gray},
  keywordstyle=\color{blue},
  keywordstyle=[2]\color{dkgreen},
  keywordstyle=[3]\color{magenta},
  commentstyle=\color{gray},
  stringstyle=\color{mauve},
  breaklines=true,
  breakatwhitespace=true,
  tabsize=3,
  frame=none,
}

\lstdefinestyle{cstyleRules}{
  escapechar={@},
  language=c,
  aboveskip=3mm,
  belowskip=3mm,
  showstringspaces=false,
  columns=flexible,
  basicstyle={\scriptsize\ttfamily},
  numbers=none,
  numberstyle=\tiny\color{gray},
  keywordstyle=\color{blue},
  keywordstyle=[2]\color{dkgreen},
  keywordstyle=[3]\color{magenta},
  commentstyle=\color{gray},
  stringstyle=\color{mauve},
  breaklines=true,
  breakatwhitespace=true,
  tabsize=3,
  keywords=[2]{cexpr,bin_oper,cop,vector_space,scalar_field,contains_expr,una_oper,cstmts,cstmt,subs,no_mod,no_mod_use,is_distributive,has_expr,no_deps,new_var,ctype,replace, pure, times},
  keywords=[3]{pattern,generate,condition,metrics,assert},  
}

\lstdefinestyle{cstyleTikz}{
  escapechar={@},
  language=c,
  aboveskip=3mm,
  belowskip=3mm,
  showstringspaces=false,
  columns=flexible,
  basicstyle={\scriptsize\ttfamily},
  numbers=none,
  numberstyle=\tiny\color{gray},
  keywordstyle=\color{blue},
  keywordstyle=[2]\color{dkgreen},
  keywordstyle=[3]\color{magenta},
  commentstyle=\color{gray},
  stringstyle=\color{mauve},
  breaklines=true,
  breakatwhitespace=true,
  tabsize=3,
  keywords=[2]{cexpr,bin_oper,cop,vector_space,scalar_field,contains_expr,una_oper,cstmts,cstmt,subs,no_mod,no_mod_use,is_distributive,has_expr,no_deps,new_var,ctype,replace, pure, times},
  keywords=[3]{pattern,generate,condition,metrics},  
}

\newcommand{\ihaskell}[1]{\lstinline[style=hstyle,basicstyle={\small\ttfamily}]{#1}}

\newcommand{\rw}{\ensuremath{\Rightarrow}}
\newcommand{\when}{\ensuremath{\mathbf{when}}}

\newcommand{\nowrites}{\centernot{\mapsto}}
\newcommand{\noreads}{\centernot{\mapsfrom}}
\newcommand{\pure}{\mathrm{pure}}
\newcommand{\fresh}{\mathrm{fresh}}
\newcommand{\writes}{\mathrm{writes}}
\newcommand{\occurs}{\mathrm{occurs~in}}

\newcommand{\distributesover}{\mathrm{distributes\_over}}
\newcommand{\nowritesexcarrays}{\underset{_{-a[l]}}{\nowrites}}
\newcommand{\nowritesarray}{\underset{_{a[l]}}{\overset{_<}{\nowrites}}}

\usepackage{soul}
\sethlcolor{pink}

\newcommand{\stml}{\textsc{stml}\xspace}

\usepackage{stmaryrd}

	\pgfmathsetmacro\lheight{1.6}
	\pgfmathsetmacro\dwidth{1.0}
	\pgfmathsetmacro\roffset{9}
	\pgfmathsetmacro\loffset{1.5}

\newcommand{\showrule}[2]{%
\begin{figure}
 \lstinputlisting[style=cstyleRules]{#2}
 \caption{\stml rule for \textsc{#1}.}
 \label{fig:stml-rule-#1}
\end{figure}
}

\begin{document}

\setlength{\pdfpageheight}{\paperheight}
\setlength{\pdfpagewidth}{\paperwidth}

\conferenceinfo{PROHA'16}{March 12, 2016, Barcelona, Spain}
\copyrightyear{2016}

\copyrightdata{978-1-nnnn-nnnn-n/yy/mm}
\copyrightdoi{nnnnnnn.nnnnnnn}

\publicationrights{transferred}
\publicationrights{licensed}     
\publicationrights{author-pays}

\preprintfooter{PROHA'16, March 12, 2016, Barcelona, Spain}   

\title{Towards a Semantics-Aware Code Transformation Toolchain for
       Heterogeneous Systems} 

\authorinfo{Salvador Tamarit \and Julio Mari\~no}
           {Universidad Polit\'ecnica de Madrid \\
             Campus de Montegancedo 28660 \\
             Boadilla del Monte, Madrid, Spain}
           {\{salvador.tamarit,julio.marino\}@upm.es}
\authorinfo{Guillermo Vigueras \and Manuel Carro}
           {IMDEA Software Institute \\
             Campus de Montegancedo 28223 \\
             Pozuelo de Alarc\'on, Madrid, Spain}
           {\{guillermo.vigueras,manuel.carro\}@imdea.org }

\maketitle

\begin{abstract}
Obtaining good performance when programming heterogeneous computing
platforms poses significant challenges for the programmer.  
We present a program transformation environment, implemented in
Haskell, where architecture-agnostic scientific C code with semantic
annotations is transformed into  functionally equivalent code better
suited for a given platform.  
The transformation steps are formalized (and implemented) as rules
which can be fired when certain syntactic and 
semantic conditions are met.  These conditions are to be fulfilled by
program properties which can 
be automatically inferred or, alternatively, stated as annotations in
the source code.   
Rule selection can be guided by heuristics derived from a machine
learning procedure
which tries to capture how run-time characteristics
(e.g., resource consumption or performance) are affected by the transformation
steps. 
\end{abstract}

{---Code generation}
         {---Heterogeneous (hybrid) systems} 
         {---Program transformation}



\section{Introduction}
There is currently a strong trend in high-performance computing
towards the integration of various types of computing elements: vector
processors, GPUs being used for non-graphical purposes, FPGA modules,
etc.\ interconnected in the same architecture.  Each of these
components is specially suited for some class of computations, which
makes the resulting platform able to excel in performance by mapping
computations to the unit best suited to execute them and is proving to be
a cost-effective alternative to more traditional supercomputing
architectures~\cite{danalis2010-shoc-benchmark}.  However, this
specialization comes at the price of additional hardware and, notably,
software complexity.  Developers must take care of very different
features to make the most of the underlying computing
infrastructure.  Thus, programming these systems is restricted to a few
experts, which hinders its widespread adoption, increases the
likelihood of bugs and greatly limits portability.

Defining programming models that ease the task of efficiently
programming heterogeneous systems is an objective of many ongoing
efforts, among them the
European research project POLCA.\footnote{%
  \emph{Programming Large Scale Heterogeneous Infrastructures},\\ 
  \url{http://polca-project.eu}.}
The project specifically targets scientific programming on   
heterogeneous platforms, due to the performance 
attained by certain hardware components for some classes of
computations -- e.g., GPUs and linear algebra -- and to the energy
savings achieved by heterogeneous computing in scientific
applications characterized by high energy
consumption~\cite{danalis2010-shoc-benchmark,lindtjorn2011-beyond-micro}.
Additionally, most scientific applications rely on a large base of
existing algorithms that must be ported to the new architectures in a
way that gets the most out of their computational strengths, while
avoiding pitfalls and bottlenecks, and preserving the meaning of the
original code.
%
Porting is carried out by transforming or replacing certain fragments
of code to improve their performance in a given architecture while
preserving their meaning.
Unfortunately, (legacy) code often does not spell its meaning or the
programmer's intentions clearly, although scientific code usually
follows patterns rooted in its mathematical origin.

\begin{figure*}[t]
 \begin{center}
   \setlength\dashlinedash{1pt}
   \setlength\dashlinegap{2pt}
   \setlength\extrarowheight{-2ex} 
   \begin{tabular}{:l:l:l:}
     \hdashline
      \multicolumn{1}{:c:}{0 - original code} &
     \multicolumn{1}{:c:}{1 - \textsc{For-LoopFusion}} &
     \multicolumn{1}{:c:}{2 - \textsc{AugAdditionAssign}}
     \\\hdashline
     \begin{minipage}[t]{4cm}
     \vspace*{-1.5ex}
\begin{lstlisting}
float c[N], v[N], a, b;
@\aftergroup\blackcolor@@\hl{\textbf{for}(\textbf{int} i=0;i<N;i++)}@
  @\hl{c[i] = a*v[i];}@

@\hl{\textbf{for}(\textbf{int} i=0;i<N;i++)}@
  @\hl{c[i] += b*v[i];}@
\end{lstlisting}
     \vspace*{-1.5ex}
     \end{minipage}
     &
     \begin{minipage}[t]{4cm}
     \vspace*{-1.5ex}
\begin{lstlisting}
@\aftergroup\redcolor@for(int i=0;i<N;i++) {
   c[i] = a*v[i];
   @\hl{c[i] += b*v[i]}@;
}@\aftergroup\blackcolor@
\end{lstlisting}
     \vspace*{-1.5ex}
     \end{minipage}
     &
     \begin{minipage}[t]{4cm}
     \vspace*{-1.5ex}
\begin{lstlisting}
@\aftergroup\blackcolor@for(int i=0;i<N;i++) {
   @\hl{c[i] = a*v[i];}@
   @\aftergroup\redcolor@@\hl{c[i] = c[i] + b*v[i]}@@\aftergroup\blackcolor@@\hl{;}@
}
\end{lstlisting}
     \vspace*{-1.5ex}

     \end{minipage}
     \\\hdashline
     \multicolumn{1}{:c:}{3 - \textsc{JoinAssignments}} &
     \multicolumn{1}{:c:}{4 - \textsc{UndoDistribute}} &
     \multicolumn{1}{:c:}{5 -\textsc{LoopInvCodeMotion}}
     \\\hdashline
     \begin{minipage}[t]{4cm}
     \vspace*{-1.5ex}
\begin{lstlisting}
for(int i=0;i<N;i++)
  @\aftergroup\redcolor@c[i] = @\hl{a*v[i]+b*v[i]}@;
\end{lstlisting}
     \vspace*{-1.5ex}
     \end{minipage}
     &
     \begin{minipage}[t]{4cm}
     \vspace*{-1.5ex}
\begin{lstlisting}
@\aftergroup\blackcolor@for(int i=0;i<N;i++)
   @\hl{c[i] = }@@\aftergroup\redcolor@@\hl{(a+b) * v[i]}@@\aftergroup\blackcolor@@\hl{;}@
\end{lstlisting}
     \vspace*{-1.5ex}
     \end{minipage}
     &
     \begin{minipage}[t]{4cm}
     \vspace*{-1.5ex}
\begin{lstlisting}
@\aftergroup\redcolor@float k = a + b;@\aftergroup\blackcolor@
for(int i=0;i<N;i++)
   @\aftergroup\redcolor@c[i] = k * v[i];@\aftergroup\blackcolor@
\end{lstlisting}
     \vspace*{-1.5ex}
     \end{minipage}
     \\\hdashline
   \end{tabular}
 \end{center}
 \caption{A sequence of transformations of a piece of C code to compute
          $\textbf{c}=a\textbf{v}+b\textbf{v}$.} 
 \label{fig:code_trans_seq}
\end{figure*}

Our goal is to obtain a framework for transformation of scientific code where the validity of a
given transformation is guided by high-level annotations expressing
the mathematical foundation of the source code. 
Despite the broad range of compilation and refactoring tools
available~\cite{Bagge03CodeBoost,visser04:stratego-XT-0.9,Schupp2002}, 
no existing tool fits the needs of the project (see Section \ref{sec:related} for further details), 
so we decided to implement our own
transformation framework, including a domain specific language for the
definition of semantically sound code transformation rules (\stml), and
a transformation engine working at \emph{abstract syntax tree} (AST) 
level.\footnote{\url{http://goo.gl/yuOFiE}}

Therefore, we have developed a program transformation environment, implemented in
Haskell, where architecture-agnostic C code 
is transformed into a functionally equivalent one better
suited for a given platform.
Haskell provides a good abstraction to represent and manipulate ASTs. Additionally, the use of pattern matching instead of visitor pattern avoids a lot of boilerplate code, making the system less error prone. 
Rules are written using a C-like syntax called \stml, inspired by CTT~\cite{Boekhold1999} and 
CML~\cite{Brown2005-tr-opt_trans_hw}, which makes it
easy for C programmers to understand their meaning and to define them,
while the rules can transparently access core functionality provided
by the Haskell rewriting engine and be accessed by it to
select rules and blocks of code where these rules can be
safely applied.
The tool is designed to use external facilities to help in selecting
which rules have to be applied; we are developing a machine
learning-based tools which work as external oracles to automate the selection of the most promising transformation chain(s)~\cite{vigueras16:learning_code_trans}. 
The tool also includes an
interactive mode to allow for more steering by expert users.
Finally, when a optimal code is reached, it is translated to adapt to the target platform programming model. 

Fig.~\ref{fig:code_trans_seq} shows a sample code
transformation sequence,
containing the original fragment of C code along with the result of
applying \emph{loop-fusion}, reorganizing assignments, algebraic
rewriting based on \emph{distributivity}, and moving \emph{invariant}
expressions out of the body loop. Some of these transformations are
currently done by existing optimizing compilers.  However, they are
usually performed internally, at the IR level, and without any
possibility for user intervention or tailoring, which falls short to
cater for many relevant situations which we want to address:

\begin{itemize}
\item Most compilers are designed to work without (or with minimal)
  human intervention, relying solely on static analysis.  While when
  this is possible the situation is optimal, in many cases static
  analysis cannot uncover the underlying properties that a programmer
  may know.  For example, in Fig.~\ref{fig:code_trans_seq} any
  compiler would rely on native knowledge of the properties of
  multiplication and addition.  However, if these operations were
  substituted by calls to implementations of operations with the same
  properties (distributivity, associativity, commutativity) such as
  operations on matrices, the transformation presented would be
  feasible but unlikely to be performed by a compiler relying solely
  on static analysis.

\item Most compilers have a set of \emph{standard} transformations
  which are useful in common cases for usual architectures --- usually
  Von Neumann--based CPU architectures.  However, when CPU-generic
  code is to be adapted for a specific architecture (e.g., FPGA,
  GPGPU) the transformations to be made prior to the implementation
  are not trivial and fall outside those usually implemented in
  standard compilers.  Even more, compilers such as
  ROCCC~\cite{roccc-manual}, which accept a subset of the C language
  and generate executables or lower-level code for a specific
  architecture, need the input code to follow specific coding
  patterns, which our tool can help generate.

\item Related to the previous point, transformations to generate code
  to be amenable to be compiled down to some hybrid architecture can
  be sometimes complex and are better expressed at a higher level
  rather than inside a compiler's architecture.  That could need users
  to come up with transformations which are better suited for a given
  coding style or application domain.  Therefore, giving programmers
  the possibility of defining transformations at a higher level and as
  plugins for a compiler greatly enlarges the set of scenarios where
  automatic program manipulation can be applied.
\end{itemize}


\begin{figure*}
 \begin{center}
\begin{tikzpicture}[
    place1/.style={circle,draw=blue!50,fill=blue!20,thick},
    place2/.style={circle,draw=blue!50,fill=green!80,thick},
    place3/.style={circle,draw=blue!50,fill=red!40,thick},
    thick,->,scale=0.6]
    \node (gpu) at (4.5,1.5) [place3,label=right:GPGPU (\textit{OpenCL}),label={[yshift=0.1cm,xshift=0.5cm]above:\parbox{7em}{\textcolor{red!80}{Translated code}}}] {};
    \node (omp) at (4.5,0.5) [place3,label=right:OpenMP] {};
    \node (mpi) at (4.5,-0.5) [place3,label=right:MPI] {};
    \node (fpga) at (4.5,-1.5) [place3,label=right:FPGA (\emph{\textit{MaxJ}, \textit{POROTO}})] {};
    \node (dsp) at (4.5,-2.5) [place3,label=right:DSP (\emph{\textit{FlexaWare}})] {};    
    \node (rgpu) at (2,1.5)
    [place2,label=above:\parbox{5em}{\textcolor{DarkGreen!80}{\emph{Ready}
        code}}] {} edge (gpu);
    \node (romp) at (2,0.5) [place2] {} edge (omp);
    \node (rmpi) at (2,-0.5) [place2] {} edge (mpi);
    \node (rfpga) at (2,-1.5) [place2] {} edge (fpga);
    \node (rdsp) at (2,-2.5) [place2] {} edge (dsp);    
    \node (original) at (-2.0,0)
    [place1,label=above:\parbox{4em}{\textcolor{blue!60}{Initial
        code}}] {} edge (rgpu) edge (romp) edge (rmpi) edge (rfpga) edge (rdsp);
    \draw[-] (-1.6,-3) -- (-1.6,-3.25) -- (1.95, -3.25) -- (1.95, -3);
    \draw[-] (2.05,-3) -- (2.05,-3.25) -- (4.6, -3.25) -- (4.6, -3);
    \node [blue] at (0,-3.75) {Transformation};
    \node [blue] at (3.3,-3.75) {Translation};
\node (gears) at (5.8,-5) {\includegraphics[width=3cm]{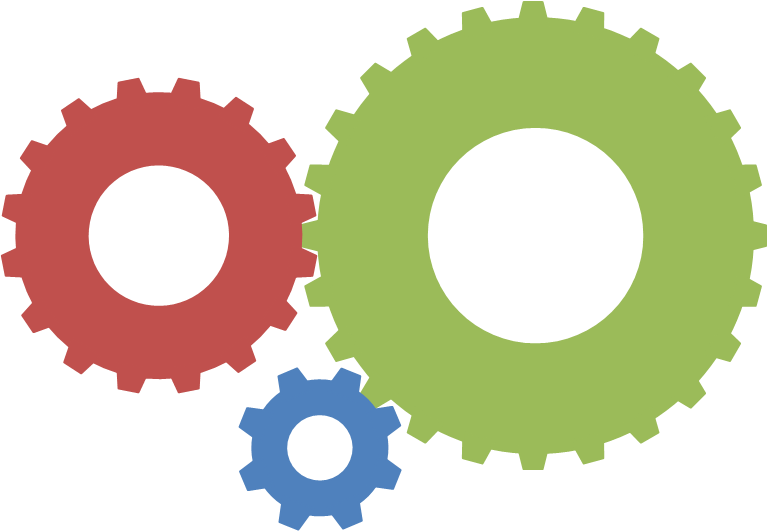}};
\node[blue,left=of gears,xshift=1.2cm] (engine) {Engine written in Haskell};
\draw[->] (-1.8, -4.4) -- (5.6, -4.4);

\node[blue,left=of original,xshift=-0.2cm,yshift=0.5cm] (rulelib)
    {\parbox{4em}{Rule  library (\stml)}};
\node[below of=rulelib,yshift=-1em] (r1)
    {\includegraphics[width=0.7cm]{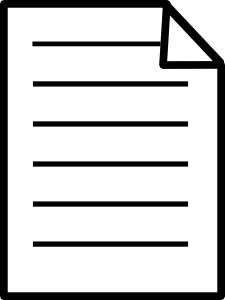}};
\node[below of=r1] (r2)
    {\includegraphics[width=0.7cm]{figures/paper}};
\node[below of=r2] (r3)
    {\includegraphics[width=0.7cm]{figures/paper}};

\node[blue,below of=engine,left=of engine,xshift=-0.5cm] (rulelibh)
    {\parbox{5em}{Rule library (Haskell)}};
\node[right=of rulelibh,xshift=-1cm] (h1)
    {\includegraphics[width=0.7cm]{figures/paper}
     \hspace*{-1.7em}\raisebox{1.2em}{\large\textbf{.hs}}};
\node[right=of h1,xshift=-1cm] (h2)
    {\includegraphics[width=0.7cm]{figures/paper}
     \hspace*{-1.7em}\raisebox{1.2em}{\large\textbf{.hs}}};
\node[right=of h2,xshift=-1cm] (h3) 
    {\includegraphics[width=0.7cm]{figures/paper}
     \hspace*{-1.7em}\raisebox{1.2em}{\large\textbf{.hs}}};

\draw[->] (r3) -- (h1);
\draw[->] (r2) -- (h2);
\draw[->] (r1) -- (h3);
\draw[->] (h1) -- (engine);
\draw[->] (h2) -- (engine);
\draw[->] (h3) -- (engine);

\node[blue] (rule) [right=of original,yshift=2.8cm,xshift=-1cm]
     {Rule execution(s)};
\path[-o] (rule.south) edge node[left] {\includegraphics[width=0.7cm]{figures/paper}} (0,0.1);
\path[->] (engine.west) edge [bend left=70]  (rule.west);
  \end{tikzpicture}
  \end{center}
  \caption{Anatomy of the transformation tool.}
  \label{fig:ana-trans-tool}
 \end{figure*}

Figure \ref{fig:ana-trans-tool} gives an overview of the tool. It is
designed to work in two stages: a \emph{transformation} phase
(Section~\ref{sec:s2stransformation}) and a \emph{translation} phase 
(Section~\ref{sec:translation}).  The former rewrites the
original input code into an intermediate format, in the same input
language, but which follows coding patterns closer to what would be
necessary to compile into the destination architecture.  The tools
implementing the transformation phase use a rule library, written in a
DSL (Sections~\ref{sec:high_level_ann} and~\ref{sec:stml_ann}), which
defines the available transformations.


The transformed code can then be fed to compilers which accept C code
adapted to the targeted architecture or be translated into code which
can then be compiled for the appropriate architecture by tools which
do not accept (sequential) C code.\footnote{For example, and in our
  case, MaxJ~\cite{techologies16:max_compiler} code can be generated or C
  code with OpenMP annotations or with MPI calls. }
%
Most of the work has currently been put into the transformation
phase. Initial work on the translation phase shows encouraging results
and point to next steps which we explain in more detail in
Section~\ref{sec:conclusions}.



\section{Related Work}
\label{sec:related}

Rule-based program transformation is a large and fruitful
area~\cite{Visser2005,van2003-prog-transf-mech}.
Our proposal has as distinguishing features the focus on scientific code and
the aim to achieve efficiency on heterogeneous
platforms, optionally using external \emph{oracles}
to guide which transformations to apply.
Also, there is a crucial distinction between systems that
generate new code from the mathematical model of an
implementation into a new model and then generating new code,
and those which use mathematical properties to transform an existing code. 
The former (automatic code synthesis) has long
been subject of research and can generate underperforming code
because of its generality.  The latter usually requires that the initial code
is in some ``canonical'' form.
Our approach, based on chaining  sound small-step code
transformations, tries to avoid those problems.
\begin{enclosedreminders}
\mclcomm{It's not clear that this makes a lot
  of sense here, but OTOH I think it's something that applies to all
  of the related work, given that many of our ideas are already
  elsewhere --- but we don't have any example with metrics, i.e., we
  don't show anything we can say is original.}
\end{enclosedreminders}
%


\begin{enclosedreminders}
  \mclcommin{Other points are how do Stratego and others deal with
  metrics (if they), properties which are decided using external
  tools, types of transformations (e.g. does it do procedural transformations
  involving destructive assignment --- but I guess the answer is
  yes).}
\mclcommin{One more thing: (semantic) annotations / hints, of course}
\end{enclosedreminders}

CodeBoost~\cite{Bagge03CodeBoost}, built on top of
Stratego-XT~\cite{visser04:stratego-XT-0.9},
performs domain-specific optimizations to C++ code
following an approach similar in spirit to our proposal.  User-defined rules specify
domain-specific optimizations; code annotations are used as
preconditions and inserted as postconditions during the rewriting
process.
\begin{enclosedreminders}
\mclcomm{Is metrics the only difference?  How about strategies?}
\end{enclosedreminders}
Concept-based frameworks such as Simplicissimus~\cite{Schupp2002}
transform C++ based on user-provided algebraic properties.
The rule application strategy can be guided by the cost of the resulting
operation, although this is done at the expression level (and not at
the statement level).
The transformation of C-like programs so as to optimize parallelism in
its compilation into a FPGA is treated in
Handel-C~\cite{Brown2005-tr-opt_trans_hw}.  It is however focused on a
synchronous language, and therefore some of its assumptions are not
valid in more general settings.
A completely different approach is to use linear algebra to transform the
mathematical specification of concrete scientific
algorithms~\cite{Franchetti2006,Fabregat2013,DiNapoli2014}.
Here, the starting point is a mathematical formula and, once the formula is
transformed, code is generated for the resulting expression. 
However, the good acceleration factors over hand-tuned code shown
happens only for those algorithms, and applying the ideas to
other contexts -- like the aforementioned reuse of legacy code --
 does not seem straightforward.


There are some language-independent transformation tools that share
some similarities with our approach. The most relevant are Stratego~\cite{visser04:stratego-XT-0.9}, TXL~\cite{Cordy2006txl}, DMS~\cite{baxter2004dms} and Rascal~\cite{klint2009rascal}. Stratego is
more oriented to strategies than to rewriting rules.  It is not
designed to add analyzers as plugins and,  additionally, it does not support
pragmas and does not keep a symbol table. The last two disadvantages are
shared with TXL. DMS is a powerful,  industrial transformation tool that is not
free and there is not too much information of how it works internally,
and its overall open documentation is scarce. It transforms programs
by applying rules until a fix point is reached, so the rules should be
carefully defined so that they do not produce loops in the rewriting
stage.  Finally, Rascal is still in alpha state and only available as
binary, so the source code is not accessible.

\section{Source-to-Source Transformations}
\label{sec:s2stransformation}

The code transformation tool has two main components: a parser for the
input language and the transformation rules written in a domain-specific language, which
translates the rules into Haskell for faster execution, and an engine to
perform source-to-source C code transformations based on these rules,
possibly making use of information provided in code annotations given
as \emph{pragmas}.

The transformation rules are written in a domain-specific language
which concisely expresses patterns which match input code and describe
the skeleton of the code to generate.  Rules state
the conditions required to soundly apply a given transformation as
well.  The annotations can capture properties at two different levels:
High-level properties which describe algorithmic structures and
low-level properties which describe details of the procedural code.
%
%
The decision of whether to apply a given transformation depends on
several factors:

\begin{itemize}
\item First, it is necessary to ensure that applying a rule at a
  certain point is sound.  Program properties to be matched with rule
  conditions can, in simple cases, be automatically verified by the
  parser.
  If this is not the case, they can be stated as annotations in the
  source code.  These annotations
  may come from external tools, such as static analyzers (to extract,
  for example, data dependency and type information), 
    or
  be provided by a programmer.
\item Second, it would be necessary to ascertain whether the
  transformation will (eventually) improve efficiency, which is far
  from trivial.  An interactive mode which makes it possible for a
  user to select which rule has to be applied is available.  However,
  the difficulty of deciding the best rule at every moment and for
  every architecture, and the overhead this imposes, makes us lean
  towards an automatic procedure.  We are developing a machine
  learning-based oracle~\cite{vigueras16:learning_code_trans} which
  decides, for every destination architecture, which rule to apply
  based on estimations of candidate transformation chains to generate
  code with the best possible performance.


\end{itemize}

We will next introduce the code annotations
and then we will continue with the rule language.  We will close this
section with a description of the interaction between the
transformation tool and the machine learning-based oracle.



\begin{table}
\centering
\begin{tabular}{|l|}
\hline   
\multicolumn{1}{|c|}{\textsc{polca annotation}}\\
\hline
\begin{lstlisting}[basicstyle = \ttfamily\scriptsize]
#pragma polca map F v w
\end{lstlisting}\\
\hline
\multicolumn{1}{|c|}{\textsc{translation to} \stml}\\
\hline  
\begin{lstlisting}[basicstyle = \ttfamily\scriptsize]
#pragma stml reads v in {0}
#pragma stml writes w in {0}
#pragma stml same_length v w
#pragma stml pure F
#pragma stml iteration_space 0 length(v)
#pragma stml iteration_independent
\end{lstlisting}\\
\hline

\multicolumn{1}{c}{~}\\

\hline 
\multicolumn{1}{|c|}{\textsc{polca annotation}}\\
\hline
\begin{lstlisting}[basicstyle = \ttfamily\scriptsize]
#pragma polca fold F INI v a
\end{lstlisting}\\
\hline
\multicolumn{1}{|c|}{\textsc{translation to} \stml}\\
\hline  
\begin{lstlisting}[basicstyle = \ttfamily\scriptsize]
#pragma stml reads v in {0} 
#pragma stml reads output(INI)
#pragma stml writes a
#pragma stml pure F
#pragma stml iteration_space 0 length(v)
\end{lstlisting}\\
\hline

\multicolumn{1}{c}{~}\\

\hline
\multicolumn{1}{|c|}{\textsc{polca annotation}}\\
\hline
\begin{lstlisting}[basicstyle = \ttfamily\scriptsize]
#pragma polca itn F INI n w
\end{lstlisting}\\
\hline
\multicolumn{1}{|c|}{\textsc{translation to} \stml}\\
\hline  
\begin{lstlisting}[basicstyle = \ttfamily\scriptsize]
#pragma stml reads output(INI)
#pragma stml reads n
#pragma stml writes w
#pragma stml pure F
#pragma stml iteration_space 0 n
\end{lstlisting}\\
\hline    

\multicolumn{1}{c}{~}\\

\hline  
\multicolumn{1}{|c|}{\textsc{polca annotation}}\\
\hline
\begin{lstlisting}[basicstyle = \ttfamily\scriptsize]
#pragma polca zipWith F v w z
\end{lstlisting}\\
\hline
\multicolumn{1}{|c|}{\textsc{translation to} \stml}\\
\hline  
\begin{lstlisting}[basicstyle = \ttfamily\scriptsize]
#pragma stml reads v in {0}
#pragma stml reads w in {0}
#pragma stml writes z in {0}
#pragma stml same_length v w
#pragma stml same_length v z
#pragma stml pure F
#pragma stml iteration_space 0 length(v)
#pragma stml iteration_independent
\end{lstlisting}\\
\hline

\multicolumn{1}{c}{~}\\

\hline  
\multicolumn{1}{|c|}{\textsc{polca annotation}}\\
\hline
\begin{lstlisting}[basicstyle = \ttfamily\scriptsize]
#pragma polca scanl F INI v w
\end{lstlisting}\\
\hline
\multicolumn{1}{|c|}{\textsc{translation to} \stml}\\
\hline  
\begin{lstlisting}[basicstyle = \ttfamily\scriptsize]
#pragma stml reads output(INI)
#pragma stml reads v in {0}
#pragma stml reads w in {0}
#pragma stml writes w in {1}
#pragma stml pure F
#pragma stml iteration_space 0 length(v)
\end{lstlisting}\\
\hline 
\end{tabular}
\caption{POLCA annotations and its translation into \stml annotations.}
\label{tab:polca_stml_translation}
\end{table}

\subsection{High-Level Annotations}
\label{sec:high_level_ann}


Information regarding the code to be transformed can be provided with
a high-level interface where annotations
describe semantic features of the code  following a functional
programming style.  This makes it possible to 
capture algorithmic skeletons at a higher level of abstraction and to
express properties of the underlying code.
For instance, \texttt{for} loops expressing a mapping between an input and
an output array can be annotated directly with a \texttt{map} pragma
such as \texttt{\#pragma polca map F v w}.  This would indicate that
the loop   traverses the input array \texttt{v}, applies function
\texttt{F} to each element in \texttt{v}, and stores the result in
\texttt{w}.  Additionally, and for the annotation to be correct, we
require that \texttt{F} is pure, that \texttt{v} and \texttt{w} have
the same length, and that every element in \texttt{w}
is computed only from the corresponding element in \texttt{v}.

The boxes labeled ``\textsc{polca annotation}'' in
Table~\ref{tab:polca_stml_translation} list the main high-level
annotations that can be currently used.
For illustrative purposes, Listing~\ref{lst:polca_ann_example} shows
an annotated version of the code in Figure~\ref{fig:code_trans_seq}.
The annotation in the second loop uses the constructor \texttt{zip} to
express that a pair of arrays is treated as an array of pairs ---that
is necessary due to the signature of \texttt{map}.
%

\begin{lstlisting}[basicstyle =
  \ttfamily\scriptsize,caption=Annotations for the initial code in Figure \ref{fig:code_trans_seq}.,label=lst:polca_ann_example]

float c[N], v[N], a, b;

#pragma polca map BODY1 v c
for(int i=0;i<N;i++)
#pragma polca def BODY1
#pragma polca input v[i]
#pragma polca output c[i]
   c[i] = a*v[i];

#pragma polca map BODY2 zip(v,c) c
for(int i=0;i<N;i++)
#pragma polca def BODY2
#pragma polca input (v[i],c[i])
#pragma polca output c[i]
   c[i] += b*v[i];
\end{lstlisting} 


\subsection{STML Properties}
\label{sec:stml_ann}

A source-to-source transformation tool for a procedural language
requires information about the code (i.e. code properties) to ensure
that transformations are correctly applied.  These properties are of a
lower level than those directly captured by high-level annotations,
because they have to deal with characteristics pertaining to a
concrete programming language.
%
%
For example, a purely functional semantics for the high-level annotations
cannot naturally capture some aspects of imperative languages such as
destructive assignment or aliasing, since it would have to follow the
referential transparency common in functional languages.
Likewise, manual memory management (explicit memory allocation and
deallocation) would be difficult to represent.
In our framework, these procedural code-level characteristics are
expressed in a language we have termed \stml (\emph{Semantic
  Transformation Meta-Language}) which is used both in the code
annotations and in the conditions of the transformation rules.



\subsubsection{Syntax and Semantics of \stml Annotations}
\label{sec:stml-annotations}


The BNF grammar for \stml annotations is shown in 
Listing~\ref{lst:stml_grammar}.  An intuitive explanation of its
semantics follows:

\begin{lstlisting}[ basicstyle=\scriptsize\ttfamily,
    float,caption= BNF grammar for \stml annotations.,label=lst:stml_grammar]

<code_prop_list> ::= "#pragma stml" <code_prop> 
    | "#pragma stml" <code_prop> <code_prop_list>

<code_prop>      ::= <exp_prop> <exp>
    | [<op>] <op_prop> <op> 
    | <mem_access> <exp> ["in" <offset_list>]
    | "write("<exp>") =" <location_list>
    | "same_length" <exp> <exp> 
    | "output("<exp>")"  | <loop_prop>

<loop_prop>      ::= "iteration_independent"
    | "iteration_space" <parameter> <parameter>

<exp_prop>       ::= "appears" | "pure" | "is_identity"
    
<op_prop>       ::= "commutative" | "associative"
    | "distributes_over"  

<mem_access>     ::= "writes" | "reads" | "rw"

<location_list>  ::= 
    "{" <c_location> {"," <c_location>} "}" 

<offset_list>    ::= "{" <INT> {"," <INT>} "}"

<exp>              ::= <C_EXP>  | <C_VAR> | <polca_var_id>

<op>               ::= <C_OP> | <C_VAR> | <polca_var_id>

<c_location>       ::= <C_VAR> | <C_VAR>("["<C_EXP>"]")+

<parameter>        ::= <c_location> | <polca_var_id> | <INT>
\end{lstlisting}

\begin{itemize}
\item \lstinline{<code_prop>} refers to the different code properties
  expressed through \stml annotations. By means of these annotations
  the user can provide information about read/write accesses to
  variables, loop properties, etc. The different options for this term
  are described below.  
  \jmcommin{examples?}

\item{\lstinline{[<exp>] <exp_prop> <exp>}:} The term
  \lstinline{<exp_prop>} indicates
  properties about code expressions of the statement that is
  immediately below the annotation. Some examples are: 

\begin{itemize}
\item{\lstinline{appears <exp>}}: There is at least one occurrence of
  \lstinline{<exp>} in the statement below.

\item{\lstinline{pure <exp>}}: Expression \lstinline{<exp>} is pure,
  i.e.\ it neither has side effects nor writes on any memory location.

\item{\lstinline{is_identity <exp>}}: Expression \lstinline{<exp>} is
  the identity element. This annotation must be preceded by
  high-level annotations that define the group or field in which
  \lstinline{<exp>} is the identity element.
\end{itemize}

\item{\lstinline{[<op>] <op_prop> <op>}:} The term
  \lstinline{<op_prop>} can be an unary or binary operator indicating
  properties about operators. Some examples are: 

\begin{itemize}
\item{\lstinline{commutative <op>}}: Operator \lstinline{<op>} has
  the commutative property: $\forall x, y .~ f(x,y) = f(y,x)$

\item{\lstinline{associative <op>}}: Operator \lstinline{<op>} has
  the associative property:
  $\forall x, y, z .~ f(f(x,y),z) = f(x,f(y,z))$

\item{\lstinline{<op> distributes_over <op>}}: The first operator
  distributes over the second operator:\\
  $\forall x, y, z .~ g(f(x,y),z) = f(g(x,z),g(y,z))$
\end{itemize}

\item{\lstinline{"write("<exp>") =" <location_list>}}: This term states
  the list of memory locations written on by expression
  \lstinline{<exp>}, where \lstinline{<location_list>} is a list of
  variables (scalar or array type) in the C code. For example,
  \lstinline|write(c = a + 3) = {c}| and%
 ~\lstinline|write(c[i++] = a +  3) = {c[i], i}| 

\item{\lstinline{<mem_access> <exp> ["in" <offset_list>]}}: The term
  \lstinline{<mem_access>} states properties about the memory accesses
  to expression \lstinline{<exp>} of the statement (or statements)
  that immediately follow.  When the
  expression \lstinline{<exp>} is an array, the list of array positions
  accessed can be referenced through \lstinline{"in" <offset_list>}
  where \lstinline{<offset_list>} is the list of array positions
  accessed either for writing or reading, depending on the access
  stated by \lstinline{<mem_access>}. Some examples are: 
 
 \begin{itemize}
\item{\lstinline{writes <exp>}}: This annotation specifies that the set
  of statements associated to the \stml
  annotation writes into location identified by \lstinline{<exp>}. 

\item{\lstinline{writes <exp> "in" <offset_list>}}: This annotation is
  similar to the previous one, but for non-scalar variables within
  loops.  It specifies that for each
  \texttt{i}-th iteration of the loop, an array identified by
  \lstinline{<exp>} is written to in the locations whose offset with
  respect to the index of the loop is contained in
  \lstinline{<offset_list>}. For example: 

\noindent
\begin{minipage}{\linewidth}
\begin{lstlisting}[style=cstyle]
#pragma stml writes c in {0}
for (i = 0; i < N; i++)
    c[i] = i*2;
    
#pragma stml writes c in {-1,0}
for (i = 1; i < N; i++){
    c[i-1] = i;
    c[i]    = c[i-1] * 2;    
}        
\end{lstlisting}
\end{minipage}

\item{\lstinline{reads <exp>}}:  The set of statements associated to
  the \stml annotation read from location \lstinline{<exp>}. 

\item{\lstinline{reads <exp> "in" <offset_list>}}: This annotation is
  similar to the previous one, but for non-scalar variables within
  loops: for the
  \texttt{i}-th iteration of the loop, the array identified by
  \lstinline{<exp>} is accessed to read locations with offset values
  contained in \lstinline{<offset_list>}. An example follows: 

\noindent
\begin{minipage}{\linewidth}
\begin{lstlisting}[style=cstyle]
#pragma stml reads c in {0}
for (i = 0; i < N; i++)
    a += c[i];
    
#pragma stml reads c in {-1,0,+1}
for (i = 0; i < N; i++)
    a += c[i-1]+c[i+1]-2*c[i];    
\end{lstlisting}
\end{minipage}

\item{\lstinline{rw <exp>}} The set of statements associated to the
  \stml annotation reads and writes from / into location
  \lstinline{<exp>}.
\item{\lstinline{rw <exp> "in" <offset_list>}} This annotation is
  similar to the previous one, but for non-scalar variables within
  loops.  The annotation specifies that for the
  \texttt{i}-th iteration of the loop, an array identified by
  \lstinline{<exp>} is accessed to read and/or write locations with
  offset values contained in \lstinline{<offset_list>}. 
\end{itemize}
 
\item{\lstinline{<loop_prop>}}: This term represents annotations
  related with loop properties 
 \begin{itemize}
 \item{\lstinline{"iteration_space" <parameter> <parameter>}} This
   annotation states the iteration space limits of the \texttt{for} loop
   associated with the annotation. An example would be:

\noindent
\begin{minipage}{\linewidth}
\begin{lstlisting}[style=cstyle]
#pragma stml iteration_space 0 N-1
for (i = 0; i < N; i++)
    c[i] = i*2;
\end{lstlisting}
\end{minipage}

\item{\lstinline{"iteration_independent"}} This annotation is used to state that there is no loop-carried dependencies in the body of the loop associated to this annotation.
\end{itemize}

\item{\lstinline{"same_length" <exp> <exp>}:} This annotation states that two arrays in the C code, given as parameters in \lstinline{ <exp>}, have the same length.

\item{\lstinline{"output("<exp>")"}:} This annotation is used to reference the output of a block of code identified by \lstinline{<exp>}.

\end{itemize}

\subsubsection{Translation from High-Level to \stml 
Annotations} 
\label{sec:polca-stml-translation}

When procedural code is decorated with high-level annotations, the
annotated code is assumed to implement the computation
expressed in the annotation. Additionally, the elements
which appear in the annotation are supposed to follow a functional
semantics, such as referential transparency.
%
%
Using this interpretation, lower-level \stml properties can be
inferred for the annotated code and used to decide which
transformations are applicable.

For example, consider a loop annotated with a \texttt{map F v w},
like the first one in Listing \ref{lst:polca_ann_example}, where 
\texttt{F} is \texttt{BODY1}, \texttt{v} is \texttt{v}, and \texttt{w} is \texttt{c}.
In this context, we assume that:

\begin{itemize}
\item \texttt{F} behaves as if it had no side effects.  It may read
  and write from/to a global variable, but it should behave as if this
  variable did not implement a state for \texttt{F}. For example, it
  may always write to a global variable and then read from it, and the
  behavior of other code should not depend on the contents of this
  variabe.
\item \texttt{v} and \texttt{w} are arrays of the same size.
\item For every element of \texttt{w}, the element in the $i$-th
  position is computed by applying \texttt{F} to the element in the
  $i$-th position of \texttt{v}.
\item The applications of \texttt{F} are not assumed to be done in any
  particular order (i.e., they can go from \texttt{v[0]} upwards to
  \texttt{v[length(v)-1]} or in the opposite direction.  Therefore
  all applications of \texttt{F} should be independent from each other.
\end{itemize}

The \stml properties inferred from some high-level annotations are shown in
Table~\ref{tab:polca_stml_translation}.  If we focus again our
attention on the translation of \texttt{map}, the \stml annotations
mean that:

\begin{itemize}
\item Iteration $i$-th reads from \texttt{v} in the position $i$-th
  (it actually reads in the set of positions \{$i+0$-th\}, since the
  set of offsets it reads from is \{0\}).
\item Iteration $i$-th ultimately writes on \texttt{w} in the position
  $i$-th (same comment as before).
\item \texttt{v} and \texttt{w} have the same length.
\item \texttt{F} behaves as if it did not have side effects.
\item \texttt{F} is applied to \texttt{v} and \texttt{w} in the
  indexes ranging from $0$ to $length(\mathtt{v})$. 
\end{itemize}


The correspondence of other high-level annotations (\texttt{foldl},
\texttt{itn}, \texttt{zipWith} and \texttt{scanl}) into \stml
annotations is shown in Table~\ref{tab:polca_stml_translation}. An extended explantation of these annotations can be found in~\cite{cluster2015}.
As in the case of the \texttt{map} annotation, the derived \stml
assertions provide properties of the procedural code which are used to
ensure that transformation rules are correctly applied.
An example is shown in Listing~\ref{lst:map_translation_example},
which corresponds to the translation into \stml annotations of the
code in Listing~\ref{lst:polca_ann_example}, which implements the
computation of a \texttt{map}.

\begin{lstlisting}[float, caption=Code from Listing \ref{lst:polca_ann_example} after translating high-level annotations to \stml.,label=lst:map_translation_example]

float c[N], v[N], a, b;

#pragma polca map BODY1 v c
#pragma stml reads v in {0}
#pragma stml writes c in {0}
#pragma stml same_length v c
#pragma stml pure BODY1
#pragma stml iteration_space 0 length(v)
#pragma stml iteration_independent
for(int i = 0; i < N; i++)
#pragma polca def BODY1
   c[i] = a*v[i];

#pragma polca map BODY2 zip(v,c) c
#pragma stml reads (v in {0}, c in {0})
#pragma stml writes c in {0}
#pragma stml same_length zip(v,c) c
#pragma stml pure BODY2
#pragma stml iteration_space 0 length(zip(v,c))
#pragma stml iteration_independent
for(int i = 0; i < N; i++)
#pragma polca def BODY2
   c[i] += b*v[i];
\end{lstlisting}

%

The \stml annotations for the code are internally used by the
source-to-source transformation tool to decide which transformations
can be applied.  For example in this case, the \textsc{For-LoopFusion}
transformation (Table~\ref{tab:transformations-math-1}) needs certain
non-syntactical properties to be met. A description of the \stml
properties used for the transformations in Figure~\ref{fig:code_trans_seq} is shown in Table \ref{tab:basic-predicates}.

%
%
%
%

\begin{table*}[t]
\begin{center}
\begin{tabular}{l@{~~~}l}
$s \nowrites\ l$ &
 statements $s$ do not write into location $l$: $l \notin \writes(s)$ \\
$s \noreads\ l$ &
 statements $s$ do not read the value in location $l$\\
$s_1 \nowrites\ s_2$ &
 statements $s_1$ do not write into any location read by $s_2$\\
 $s_1 \noreads\ s_2$ &
 statements $s_1$ do not read from any location written by $s_2$\\
 $s_1 \nowritesexcarrays\ s_2$ &
 same predicate than previous one but not taking into account locations referred through arrays.\\
 $s_1 \nowritesarray\ s_2$ &
 statements $s_1$ do not write into any previous location corresponding to an index array read by $s_2$\\
$e~\pure$ &
 expression $e$ is \emph{pure}, i.e.~does not have side effects nor
 writes any memory locations.\\
 $\writes(s)$ &
 set of locations written by statements $s$.\\
{$g~\distributesover~f$}&
$\forall x, y, z .~ g(f(x,y),z) \approx f(g(x,z),g(y,z))$\\
$l~\fresh$ &
 $l$ is the location of a \emph{fresh} identifier, i.e.~does not clash with existing
 identifiers if introduced in\\& a given program state.\\
\end{tabular}
\end{center}
\caption{Predicates used to express conditions for the application
         transformation rules in Table \ref{tab:transformations-math-1}.}
\label{tab:basic-predicates}
\end{table*}

\begin{table*}[t]
   \begin{center}
     \begin{tabular}{l@{}}
           \begin{minipage}{4,8cm}
       \begin{lstlisting}
 for($l$=$e_{ini}$;$rel(l,e_{end})$;$mod(l)$)
 {$s_1$}
 for($l$=$e_{ini}$;$rel(l,e_{end})$;$mod(l)$)
 {$s_2$}
 \end{lstlisting}
      \end{minipage}
       \rw~~
       \begin{minipage}{5.2cm}
       \begin{lstlisting}
 for($l$=$e_{ini}$;$rel(l,e_{end})$;$mod(l)$) 
 {$s_1$;$s_2$}
       \end{lstlisting}
       \end{minipage}\vspace{-2ex}\\
 	\when\ $rel~\pure$,
       \lstinline|$(s_1$;$s_2) \nowrites \{l, e_{ini}, e_{end}\}$|, $\writes(mod(l)) \subseteq \{l\},$
       $~ s_1 \nowritesexcarrays ~ s_2, s_2 \nowritesexcarrays ~ s_1, s_2 \nowritesarray ~ s_1$\\
       \hfill \textsc{(For-LoopFusion)}\\
       \lstinline|$l$ += $e$;|
        \rw\ 
       \lstinline|$l$ = $l$ + $e$;| \\ 
       \when\ $l~\pure$\\
       \hfill \textsc{(AugAdditionAssign)}\\
       \lstinline|$s_1$; $l$ = $e_1$; $s_2$; $l$ = $e_2$; $s_3$;| \rw\ 
       \lstinline|$s_1$; $s_2$; $l$ = $e_2[e_1/l]$; $s_3$;| \\ 
       \when\ $l, e_1~\pure, s_2 \nowrites\ \{l, e_1\}, s_2 \noreads\ l, s_2 \nowrites\ e_1$\\
      \hfill \textsc{(JoinAssignments)}\\
      $f(g(e_1,e_3),g(e_2,e_3))$ \rw\ $g(f(e_1,e_2),e_3)$ \\ 
       \when\ $e_1, e_2, e_3~\pure, g~\distributesover\ f$\\
       \hfill \textsc{(UndoDistribute)}\\
       \lstinline|for ($e_1$;$e_2$;$e_3$){$s_b$}| \rw\
       \lstinline|$l$ = $e_{inv}$; for ($e_1$;$e_2$;$e_3$){$s_b[l/e_{inv}])$}|\\
       \when\ $l~\fresh, e_{inv} ~\occurs~ s_b, e_{inv}~\pure, \{s_b, e_3, e_2 \}\nowrites e_{inv}$\\
       \hfill \textsc{(LoopInvCodeMotion)}\\
    \end{tabular}
  \end{center}
   \caption{Source code transformations used in the
     example of Figure \ref{fig:code_trans_seq}.}
   \label{tab:transformations-math-1}
\end{table*}

\subsubsection{Using External Tools}

Besides the properties provided by the user, either written directly
in \stml or deduced from high-level annotations, additional properties
can be obtained from external tools.  This is useful as it would
relieve users from having to write a large number of annotations which
state many low-level details.  These properties can be made available
to the transformation tool by writing them as \stml annotations.


We are currently using Cetus~\cite{dave2009cetus} to automatically
produce \stml annotations.  Cetus is a compiler framework, written in
Java, to implement source-to-source transformations. We have modified
it (which is allowed by its license) to add some new analyses and to
output the properties it infers as \stml pragmas annotating the
input code.


If the annotations automatically inferred by external tools contradict
those provided by the user, the properties provided by the user are
preferred to those deduced from external tools, but a warning is
issued nonetheless.



\subsection{STML Rules}
\label{sec:stml_rules}

\showrule{JoinAssignments}{cml_rules/rule3.c}

\begin{table*}[t]
  \begin{center}
    \begin{tabular}{| l  l |}
      \hline
      \textbf{Function/Construction} & \textbf{Description}\\
      \hline
      \texttt{subs((S|[S]|E),E$_\mathtt{f}$,E$_\mathtt{t}$)} 
      	& Replace each occurrence of \texttt{E$_\mathtt{f}$} in
        \texttt{(S|[S]|E)} for \texttt{E$_\mathtt{t}$}.\\
      \hline
      \texttt{if\_then:\{E$_\mathtt{cond}$;} 
     \texttt{(S|[S]|E);\}}
       & If \texttt{E$_\mathtt{cond}$} is true,
        ~then generate \texttt{(S|[S]|E).}\\
     \hline
   \texttt{if\_then\_else:}\texttt{\{E$_\mathtt{cond}$;}&If \texttt{E$_\mathtt{cond}$} is true, 
then generate \texttt{(S|[S]|E)$_\mathtt{t}$}\\		
\texttt{~~(S|[S]|E)$_\mathtt{t}$;}\texttt{(S|[S]|E)$_\mathtt{e}$;\}} &
\hspace*{6.2em}else generate  \texttt{(S|[S]|E)$_\mathtt{e}$}.\\
      \hline
      \texttt{gen\_list:}  \texttt{\{[(S|[S]|E)];\}}
      	&  Each element in \texttt{[(S|[S]|E)]} 
       produces a different rule consequent.\\
      \hline
    \end{tabular}
  \end{center}
  \caption{Rule language constructions and functions for \texttt{generate} rule section.}
  \label{tab:funsgenerate}
\end{table*}

\begin{table*}[t]
  \begin{center}
    \begin{tabular}{|  l  l  |}
      \hline
      \textbf{Function} & \textbf{Description}\\
      \hline
      \texttt{no\_write((S|[S]|E)$_\mathtt{1}$,} \texttt{(S|[S]|E)$_\mathtt{2}$)} 
      	&  True if \texttt{(S|[S]|E)$_\mathtt{1}$} does not write
      ~in any location read by \texttt{(S|[S]|E)$_\mathtt{2}$}.\\
      \hline
     \texttt{no\_write\_except\_arrays} 
      	&  Same than previous, but not taking into
      ~arrays accessed using \texttt{E}.\\
\texttt{~~((S|[S]|E)$_\mathtt{1}$,(S|[S]|E)$_\mathtt{2}$,}\texttt{E)}        & ~\\
      \hline
     \texttt{no\_write\_prev\_arrays} 
      	&  True if all array writes indexed using \texttt{E} in \texttt{(S|[S]|E)$_\mathtt{1}$} do not access\\ 
       \texttt{~~((S|[S]|E)$_\mathtt{1}$} \texttt{(S|[S]|E)$_\mathtt{2}$,} \texttt{E)}&  
         previous locations to array reads
        ~indexed using  \texttt{E} in \texttt{(S|[S]|E)$_\mathtt{2}$}.\\
      \hline
    \texttt{no\_read((S|[S]|E)$_\mathtt{1}$,} \texttt{(S|[S]|E)$_\mathtt{2}$)} 
      	&  True if \texttt{(S|[S]|E)$_\mathtt{1}$} does not read in
     
      ~any location written by \texttt{(S|[S]|E)$_\mathtt{2}$}.\\
      \hline
      \texttt{pure((S|[S]|E))} 
      	&  True if \texttt{(S|[S]|E)} does not write
	in any location.\\
      \hline
     \texttt{writes((S|[S]|E))} 
      	&  Locations written by \texttt{(S|[S]|E)}.\\ 
      \hline
      \texttt{distributes\_over(E$_\mathtt{1}$,E$_\mathtt{2}$)} 
      	&  True if operation \texttt{E$_\mathtt{1}$} distributes over operation \texttt{E$_\mathtt{2}$}.\\ 
      \hline
      \texttt{occurs\_in(E,(S|[S]|E))} 
      	&  True if expression \texttt{E} occurs in \texttt{(S|[S]|E)}.\\ 
      \hline
      \texttt{fresh\_var(E)} 
      	&  Indicates that \texttt{E} should be fresh. \\ 
      \hline
      \texttt{is\_identity(E)} 
      	&  True if \texttt{E} is the identity.\\ 
     \hline
      \texttt{is\_assignment(E)} 
      	&  True if \texttt{E} is an assignment.\\ 
      \hline
      \texttt{is\_subseteq(E$_\mathtt{1}$,E$_\mathtt{2}$)} 
      	&  True if \texttt{E$_\mathtt{1}$} $\subseteq$ \texttt{E$_\mathtt{2}$}
	\\ 
      \hline
    \end{tabular}
  \end{center}
  \caption{Rule language functions for the section \texttt{condition}
    of a rule.}
  \label{tab:funscondition}
\end{table*}

As an example of a rewriting rule written in \stml, let us consider
the code for the rule \textsc{JoinAssignments} shown in
Figure~\ref{fig:stml-rule-JoinAssignments}.  The rule is applied when
the program / library that we are transforming has a  piece of code matching the
pattern in the \texttt{pattern} section and it fulfills the conditions
in the \texttt{con\-di\-tion} section. The resulting code corresponds
to the pattern in the \texttt{generate} section, where expressions
matched in the \texttt{pattern} are replaced in the \texttt{generate}d
code.  In this case one assignment is removed by propagating the
expression in the assignment's right hand side.

\stml uses metavariables to match components of the initial code.
These metavariables are \emph{tagged} to denote explicitly which kind
of component they should match.  For example, a metavariable \texttt{v}
can be tagged \texttt{cexpr(v)} to denote that it can only match an
expression, \texttt{cstmt(v)} to denote that it can only match a
statement, or \texttt{cstmts(v)} to denote that it can match a sequence
of statements.  
%
In the example in Figure~\ref{fig:stml-rule-JoinAssignments},
\texttt{s1}, \texttt{s2} and \texttt{s3} should be (sequences of)
statements, and \texttt{e1}, \texttt{e2} and \texttt{l} are expressions.

There are additional conditions and primitives to help in generating
code; these 
are presented and briefly described in
Tables~\ref{tab:funscondition} and~\ref{tab:funsgenerate},
respectively.
In these tables, \texttt{E} represents an expression, \texttt{S}
represents a statement, while \texttt{[S]} represents a sequence of
statements. Additionally, the function
\texttt{bin\_oper(E$_\mathtt{op}$,E$_\mathtt{l}$,E$_\mathtt{r}$)}
matches or generates a binary operation \texttt{(E$_\mathtt{l}$
  E$_\mathtt{op}$ E$_\mathtt{r}$)} and can be used in the sections
\texttt{pattern} and \texttt{generate}.  The section \texttt{generate}
can also state, using \texttt{\#pragma}s, new properties which hold in
the resulting code.

\subsection{Rule Selection}
\label{sec:trans_ml_inter}

Our code transformation tool performs chaining of rule applications which go from
initial code to a final form where some conditions are met ---see,
e.g., Figure~\ref{fig:code_trans_seq}.  In general, several rules can
be applied at multiple code points in the rewriting process.  Deciding whether
a given rule is applicable or not depends on whether rule conditions are met or not.  In practice, it is often the case that
the tool does not have information enough to decide whether a property
holds or not, and therefore it cannot be decided whether 
%
%
a rule which requires that property can be applied or not. Thus,
the system distinguishes between definitely ``applicable'',
``definitely not applicable'' and ``probably (not) applicable''
transformation steps.  For all applicable rules, deciding which one
should be chosen
should be based on whether that rule can 
contribute to an eventual improvement of the performance of the final code with respect to the original one. 
%
We provide two ways to perform rule selection: the possibility of
having interaction with a human and a generic interface to
communicate with external tools.

%


\subsubsection{Interactive Rule Selection}
\label{sec:interactive-selection}

Our tool provides an interface to make it possible an interactive
transformation: the user is presented with the possible rules to be
applied to some code and the code before and after applying the
rule.  This can be 
useful to refine rules or 
to perform refactoring with general purposes (e.g., improve
  readability, make changes which impact large sections of code,
  perform repetitive but delicate maintenance, \ldots) which may not be
  related to improving performance or to adapt existing code to a
  given platform.
Auxiliary programs can be used to show the differences in the code before
and after applying a given transformation step (for example,
Meld~\cite{meld}) that can help the user in this process.

\subsubsection{Oracle-Based Rule Selection}
\label{sec:oracle-selection}

Due to its fine-grained nature, manual rule selection is in general not scalable: in our experience,
it is not a realistic possibility for even small programs with a
reduced set of rules.  Therefore, mechanizing as much as possible this
process is a must.
A straightforward possibility is to select at each step the
rule\footnote{Or one of them, should there be several candidates.}
which reduces some metric.  This may however make the transformation
be trapped in local minima, and in our experience, in many cases it is
necessary to apply transformations which temporarily reduce the
quality of the code because they enable the application of further transformations.

A possibility to work around this problem is to explore a bounded
number of possible rule applications ``in the vicinity of'' some
state.  How many steps should be taken in this exploration is
something which needs to be decided, since taking too few steps would
not make it possible to leave a local minima.  Given that in our
experience the number of rules that can be applied in most states is
high (typically in the order of the tens), increasing the diameter of
the boundary to be explored can cause an exponential explosion in the
number of states to be evaluated, even for such a bounded search.
 
%

Additionally, and since  rules are a parameter for the
transformation engine, the user can introduce rules which are the
inverse of each other (which may lead to infinite loops) or which
duplicate the code, such as loop unfolding rules, making it possible
to apply repeatedly the same rule to states which are different (e.g., to
different code configurations).



Among the possibilities to select a rule which is part of a promising
sequence which leads to good code for a given platform (see
Section~\ref{sec:translation}), we are exploring the use of machine
learning techniques based on 
%
\emph{reinforcement learning}~\cite{vigueras16:learning_code_trans}.
From the point of view of the transformation engine, the selection
tool works as an \emph{oracle} which, given a code configuration and a
set of applicable rules, returns which rule should be applied.  We
will describe now an abstract interface to an external rule
selector, which can be applied not only to the current oracle, but
to other similar tools.   



\newcommand{\AppRules}{\mathit{AppRules}}
\newcommand{\Code}{\mathit{Code}}
\newcommand{\Rule}{\mathit{Rule}}
\newcommand{\Pos}{\mathit{Pos}}
\newcommand{\Trans}{\mathit{Trans}}

The interface of the transformation tool is described by functions

\begin{itemize}
\item $\AppRules(\Code) \rightarrow \{(\Rule, \Pos)\}$
\item $\Trans(\Code_i, \Rule, \Pos) \rightarrow \Code_o$
\end{itemize}

The function $\AppRules$ determines the possible transformations
applicable to a given code. It returns, for a given input $\Code$, a
set of tuples containing each of a rule name $\Rule$ and the code position $\Pos$
where it can be applied (e.g., the identifier of a node in the AST). On the other hand, function $\Trans$ is
basically the application of a given transformation step. For an input
code $\Code_i$, a rule name $\Rule$, and a code position $\Pos$, it
returns the resulting code $\Code_o$ after applying the
transformation.

The interface of the reinforcement learning tool includes the
functionality described at the beginning of this section -- rule
selection and stop condition:  

\newcommand{\SelectRule}{\mathit{SelectRule}}
\newcommand{\IsFinal}{\mathit{IsFinal}}
\newcommand{\Boolean}{\mathit{Boolean}}

\begin{itemize}
\item $\SelectRule(\{(\Code_i, \{\Rule_i\})\}) \rightarrow (\Code_o, \Rule_o)$
\item $\IsFinal(\Code) \rightarrow \Boolean$
\end{itemize}

The function $\SelectRule$ selects which code has to be transformed and
which rule has to be applied to perform the transformation.  Given a set of tuples containing a code
$\Code_i$ and a set of rules $\{\Rule_i\}$, it returns a tuple
containing the chosen code $\Code_o$ and the rule $\Rule_o$ that
should be applied to $\Code_o$. Likewise, function $\IsFinal$ is used
to know whether a given code $\Code$ is considered ready for translation or not.

We are now in a position  to introduce the function that defines the
interaction between the transformation and the external oracle:

\newcommand{\NewCode}{\mathit{NewCode}}
\newcommand{\rules}{\mathit{rules}}

\begin{itemize}
\item \textbf{Header}: $\NewCode(\Code_i, \{\Rule_i\}) \rightarrow
  (\Code_o, \Rule_o)$ 
\item \textbf{Definition}: $\NewCode(c, \rules) = $\\
$\SelectRule($\\
$~~~\{(c', \{r'~|~(r',\_) \in \AppRules(c')\})$\\
$~~~ ~~|~ c' \in \{\Trans(c, r, p)$\\
$~~~ ~~~ ~~~ ~~~ ~~~ ~~|~ (r, p) \in \AppRules(c), r \in \rules \} ~\})$\\
\end{itemize} 

The function $\NewCode$ receives an initial code $\Code_i$ and a set of
rules $\{\Rule_i\}$ which are candidates to be applied. It returns the code
$\Code_o$ resulting from the application of one of the rules from
$\{\Rule_i\}$ to $\Code_i$.  Additionally, it returns the rule
$\Rule_o$ that should be applied in the next transformation step,
i.e., the next time $\NewCode$ is invoked with $\Code_o$.  The rationale behind this
design is that the first invocation receives all the applicable rules
as candidates to be applied, but after this first application we
always have a single rule in the set $\{\Rule_i\}$. We do it
repeatedly until the transformation generates a code for which function $\IsFinal$ returns true:

\newcommand{\AllRules}{\mathit{AllRules}}

\begin{itemize}
\item \textbf{Complete derivation}:\\
$\NewCode(c_0, \AllRules) \rightarrow^* (c_n,r_n)$\\
  where $\IsFinal(c_n)$ and $\forall~ 0 < i < n.\\(c_i, r_i) =
        \NewCode(c_{i -1}, \{r_{i -1}\})$ with $\neg \IsFinal(c_i)$\\\\
%
\end{itemize} 

This approach makes it unnecessary for the external oracle to have to
consider positions where a transformation can be applied, since that
choice is implicit in the selection of a candidate code between all
possible code versions obtained using a single input rule. Furthermore,
by selecting the next rule to be applied, it takes the control of the
next step of the transformation. The key here is the function
$\SelectRule$, used in function $\NewCode$.  Given an input
code $\Code_i$ and a rule $\Rule_i$,\footnote{Note that the corresponding
parameter  is a set only for the initial call.} $\SelectRule$ selects
a resulting code between all the codes that can be 
generated from $Code_i$ using $Rule_i$. The size of the set received
by function $SelectRule$ corresponds to the total number of positions
where $Rule_i$ can be applied. In this way, $SelectRule$ is
implicitelly selecting a position.

\section{Producing Code for Heterogeneous Systems}
\label{sec:translation}

In the second phase of the tool (Figure~\ref{fig:code_trans_seq}),
code for a given platform is produced starting from the result of the
transformation process.  The destination platform of a fragment of
code can be specified using annotations like this:

\medskip

\texttt{\#pragma polca mpi}

\medskip

This information is relevant, on one hand, to hint at what
transformations should be applied and also to decide when the code is
ready for translation.  In fact, the decisions taken by the machine
learning-based tool we mentioned before 
is partly directed by the destination architecture. 


The translation to a final code for a given architecture is in most
cases straightforward as it needs only to introduce the ``idioms''
necessary for the architecture or to perform a syntactical
translation.  As a consequence, there is no search or decision
process: for each input code given to the translation, there is only
one output code
which is obtained via predefined transformations or glue code
injection.

%

Some of the translations need specific information: for instance,
knowing if a statement is performing I/O is necessary when translating
to MPI, because  executing this operation might need to be done in a
single thread.  It is often the case that this can be deduced by
syntactical inspection, but in other cases (e.g., if the operation is
part of a library function) it may need explicit annotations.



\section{Implementation Notes}
\label{sec:implementation}


The transformation phase, which obtains C code that could be easily
translated into the source language for the destination platform, is a
key part of the tool.  As a large part of the system was experimental
(including the definition of the language, the properties, the
generation of the final code, and the search / rule selection
procedures), we needed a flexible and expressive implementation
platform.

We initially tested well-known infrastructures such as Clang / LLVM.
While they are very well supported and tested, we found that they
understandably were primarily designed for compilation instead of for
source-to-source program transformation, which is our main goal.  When
implementing complex source-to-source program transformation routines
in Clang, we found that the interface offered was not really designed
to perform AST transformations, and that the design documents warned
that the interface could not be assumed to be stable.  Additionally,
the methods to transform source code had to be coded in C++ which made
them verbose and full of low-level details which we did not want to
deal with.  Compiling rules to C++ was of course an option, but even
this compilation was not going to be easy, due to the conceptual
distance between the nature of the rules and the API for code
manipulation, and it was dependant upon an unstable interface.  Even
in that case, the whole Clang would have to be recompiled after
introducing new rules, which made project development and testing
cumbersome, and would make the addition user-defined rules complicated.

Therefore we decided to switch to a declarative language and implement
the tool in Haskell.  Parsing the input code is done by means of the 
\texttt{Language.C}~\cite{LanguageC} library, which returns the AST as
a data structure which is easy to manipulate.  In particular, we used
the Haskell facilities to deal with generic data structures through
the \emph{Scrap Your Boilerplate} (SYB) library~\cite{DataGenerics}.  This
allows us to easily extract information from the AST or modify it with
a generic traversal of the whole structure. 

The rules themselves are written in a subset of C, and can therefore
be also parsed using \texttt{Language.C}.  After reading them in, they
are automatically compiled into Haskell code (contained in the file
\texttt{Rules.hs} ---see Figure~\ref{fig:ana-trans-tool}) which
performs the traversal and (when applicable) the transformation of the
AST.  This module is loaded together with the rest of the tool,
therefore avoiding the extra overhead of interpreting the rules.  The
declarative nature of Haskell and facilities such as completely
automatic memory management makes this compilation much easier than it
would be in the case of compiling into C.

When it comes to rule compilation, \stml rules can be divided into two
classes: those which operate at the expression level (which are easier
to implement) and those which can manipulate both expressions and
(sequences of) statements.  In the latter case, sequences of
statements (\texttt{cstmts}) of an unknown size have to be considered:
for example, in Figure~\ref{fig:stml-rule-JoinAssignments},
\texttt{s1}, \texttt{s2}, and \texttt{s3} can be sequences of any
number of statements (including the empty sequence), and the rule has
to try all the possibilities to determine if there is a match which
meets the rule conditions. For this, Haskell code that explicitly
performs an AST traversal needs to be generated.  In the case of
expressions, they are syntactically bound, and the translation of the
rule is much easier.

When generating Haskell code, the rule sections (\texttt{pattern},
\texttt{condition}, \texttt{generate}, \texttt{assert}) generate the 
corresponding LHS's, guards, and RHS's of a Haskell function.
If the conditions to apply a rule are met, the result is returned in a
triplet \ihaskell{(rule\_name, old\_code, new\_code)} where the two
last components are, respectively, the matched and transformed
sections of the AST.  Note that \ihaskell{new\_code} may contain new
properties if the \texttt{generate} section of the rule defines them.

Since several rules can be applied at several locations of the AST,
every rewriting step can actually return a list of tuples --- one for
each rule
and location where that rule can be applied.  As mentioned elsewhere
(Section~\ref{sec:oracle-selection}), besides the possibility of
interacting with a user, we are studying the usage of an external oracle
which 
determines the best candidate to apply in the next step.  The
transformation halts when either no more rules are
applicable or when a stop condition is found, according to the
oracle. 

The tool is divided into four main modules: 

\begin{itemize}

\item \texttt{\textbf{Main.hs}} 
  implements the main  workflow of the tool: it calls the parser on the input C
  code to build the AST, links the pragmas to the AST, executes the 
  transformation sequence (interactive or automatically)
  and outputs the transformed code.

\item \texttt{\textbf{PragmaLib.hs}} reads pragmas and links them to
  their corresponding node in the AST.  It also restores or injects
  pragmas in the transformed code.

\item \texttt{\textbf{Rul2Has.hs}} translates \stml rules (stored in
  an external file) into Haskell functions which actually perform the
  AST manipulation.  It also reads and loads \stml rules as an AST and
  generates the corresponding Haskell code in the
  \texttt{\textbf{Rules.hs}} file.


\item \texttt{\textbf{RulesLib.hs}} contains supporting code used by
  \texttt{Rules.hs} to identify whether some \stml rule is or not
  applicable (e.g., there is matching code, the preconditions hold,
  etc.) and to execute the implementation of the rule (including AST
  traversal, transformation, \ldots).
\end{itemize}



\section{Conclusion}
\label{sec:conclusions}

We have presented a transformation toolchain that uses semantic
information, in the form of user- or machine-provided annotations, to
produce code for different platforms.  It has two clearly separated
phases: a source-to-source transformation which generates code with
the style appropriate for the destination architecture and a
translation from that code to the one used in the specific platform.

We have focused until now in the initial phase, which included the
specification of a DSL (\stml) to define rules and code properties, a
translator from this language into Haskell, a complete engine to work with
these rules, and an interface to interact with external oracles (such
as a reinforcement learning tool which we are developing) to guide the
transformation.

The translation phase is still in an preliminary stage. However, and
while it is able to translate some input code, it needs to be improved
in order to support a wider range of input code.  
We have compared, using several metrics, the code obtained using our
tool and the corresponding initial code and the results are
encouraging. 

As future work, we plan to improve the usability of the \stml
language.
At the same time, we are modifying Cetus to automatically obtain more
advanced\,/\,specific properties, and we are integrating profiling
techniques in the process to make it easier to evaluate the whole
transformation system and give feedback on it.

Simultaneously, we are investigating other analysis tools which can
be used to derive more precise properties.  Many of these
properties are related to data dependencies and pointer behavior.  We
are considering, on one hand, tools like
PLuTo~\cite{Bondhugula:2008} and PET~\cite{poly_pet} (two polytope
model-based analysis tools) or the dependency analyzers for the
Clang\,/\,LLVM compiler.  However, since they fall short to derive
dependencies (e.g., alias analysis) in code with pointers, we are also
considering tools based on separation
logic~\cite{DBLP:conf/csl/OHearnRY01,DBLP:conf/lics/Reynolds02} such
as VeriFast~\cite{Jacobs08theverifast,DBLP:conf/nfm/JacobsSPVPP11}
which can reason on dynamically-allocated, mutable structures.



\acks

Work partially funded by EU FP7-ICT-2013.3.4 project 610686 POLCA, Comunidad de Madrid project S2013/ICE-2731 N-Greens Software, and MINECO Projects TIN2012-39391-C04-03 / TIN2012-39391-C04-04 (StrongSoft), TIN2013-44742-C4-1-R (CAVI-ROSE), and TIN2015-67522-C3-1-R (TRACES).


\bibliographystyle{abbrvnat}
\softraggedright
\bibliography{../../BiBTeX/hpc_transformations,%
              ../../BiBTeX/polca_refs,%
              ../../BiBTeX/polca_deliverables,%
              ../../BiBTeX/c_a_t}

\begin{thebibliography}{29}
\providecommand{\natexlab}[1]{#1}
\providecommand{\url}[1]{\texttt{#1}}
\expandafter\ifx\csname urlstyle\endcsname\relax
  \providecommand{\doi}[1]{doi: #1}\else
  \providecommand{\doi}{doi: \begingroup \urlstyle{rm}\Url}\fi

\bibitem[Bagge et~al.(2003)Bagge, Kalleberg, Visser, and
  Haveraaen]{Bagge03CodeBoost}
O.~S. Bagge, K.~T. Kalleberg, E.~Visser, and M.~Haveraaen.
\newblock {Design of the CodeBoost Transformation System for Domain-Specific
  Optimisation of C++ Programs}.
\newblock In \emph{Third International Workshop on Source Code Analysis and
  Manipulation (SCAM 2003}, pages 65--75. IEEE, 2003.

\bibitem[Baxter et~al.(2004)Baxter, Pidgeon, and Mehlich]{baxter2004dms}
I.~D. Baxter, C.~Pidgeon, and M.~Mehlich.
\newblock {DMS}{\textregistered}: Program transformations for practical
  scalable software evolution.
\newblock In \emph{Proceedings of the 26th International Conference on Software
  Engineering}, pages 625--634. IEEE Computer Society, 2004.

\bibitem[Boekhold et~al.(1999)Boekhold, Karkowski, and Corporaal]{Boekhold1999}
M.~Boekhold, I.~Karkowski, and H.~Corporaal.
\newblock Transforming and parallelizing {ANSI C} programs using pattern
  recognition.
\newblock In \emph{High-Performance Computing and Networking}, pages 673--682.
  Springer, 1999.

\bibitem[Bondhugula et~al.(2008)Bondhugula, Hartono, Ramanujam, and
  Sadayappan]{Bondhugula:2008}
U.~Bondhugula, A.~Hartono, J.~Ramanujam, and P.~Sadayappan.
\newblock A practical automatic polyhedral parallelizer and locality optimizer.
\newblock \emph{SIGPLAN Not.}, 43\penalty0 (6):\penalty0 101--113, June 2008.
\newblock ISSN 0362-1340.
\newblock \doi{10.1145/1379022.1375595}.
\newblock URL \url{http://doi.acm.org/10.1145/1379022.1375595}.

\bibitem[Brown et~al.(2005)Brown, Luk, and Kelly]{Brown2005-tr-opt_trans_hw}
A.~Brown, W.~Luk, and P.~Kelly.
\newblock {Optimising Transformations for Hardware Compilation}.
\newblock Technical report, Imperial College London, 2005.

\bibitem[Cordy(2006)]{Cordy2006txl}
J.~R. Cordy.
\newblock The {TXL} source transformation language.
\newblock \emph{Science of Computer Programming}, 61\penalty0 (3):\penalty0
  190--210, 2006.

\bibitem[Danalis et~al.(2010)Danalis, Marin, McCurdy, Meredith, Roth, Spafford,
  Tipparaju, and Vetter]{danalis2010-shoc-benchmark}
A.~Danalis, G.~Marin, C.~McCurdy, J.~S. Meredith, P.~C. Roth, K.~Spafford,
  V.~Tipparaju, and J.~S. Vetter.
\newblock {The Scalable Heterogeneous Computing ({SHOC}) Benchmark Suite}.
\newblock In \emph{Proceedings of the 3rd Workshop on General-Purpose
  Computation on Graphics Processing Units}, pages 63--74. ACM, 2010.

\bibitem[Dave et~al.(2009)Dave, Bae, Min, Lee, Eigenmann, and
  Midkiff]{dave2009cetus}
C.~Dave, H.~Bae, S.~Min, S.~Lee, R.~Eigenmann, and S.~P. Midkiff.
\newblock Cetus: {A} source-to-source compiler infrastructure for multicores.
\newblock \emph{{IEEE} Computer}, 42\penalty0 (11):\penalty0 36--42, 2009.
\newblock \doi{10.1109/MC.2009.385}.
\newblock URL \url{http://dx.doi.org/10.1109/MC.2009.385}.

\bibitem[{Di Napoli} et~al.(2014){Di Napoli}, Fabregat-Traver, Quintana-Orti,
  and Bientinesi]{DiNapoli2014}
E.~{Di Napoli}, D.~Fabregat-Traver, G.~Quintana-Orti, and P.~Bientinesi.
\newblock Towards an efficient use of the {BLAS} library for multilinear tensor
  contractions.
\newblock \emph{Applied Mathematics and Computation}, 235:\penalty0 454--468,
  May 2014.

\bibitem[Fabregat-Traver and Bientinesi(2013)]{Fabregat2013}
D.~Fabregat-Traver and P.~Bientinesi.
\newblock Application-tailored linear algebra algorithms: A search-based
  approach.
\newblock \emph{International Journal of High Performance Computing
  Applications (IJHPCA)}, 27\penalty0 (4):\penalty0 425 -- 438, Nov. 2013.

\bibitem[Franchetti et~al.(2006)Franchetti, Voronenko, and
  P{\"u}schel]{Franchetti2006}
F.~Franchetti, Y.~Voronenko, and M.~P{\"u}schel.
\newblock {{FFT} Program Generation for Shared Memory: {SMP} and Multicore}.
\newblock In \emph{Supercomputing (SC)}, 2006.

\bibitem[Huber(2014)]{LanguageC}
B.~Huber.
\newblock {The Language.C Package}.
\newblock \url{https://hackage.haskell.org/package/language-c}, 2014.

\bibitem[Jacobs and Piessens(2008)]{Jacobs08theverifast}
B.~Jacobs and F.~Piessens.
\newblock The {VeriFast} program verifier.
\newblock Technical report, Department of Computer Science, KU Leuven, 2008.

\bibitem[Jacobs et~al.(2011)Jacobs, Smans, Philippaerts, Vogels, Penninckx, and
  Piessens]{DBLP:conf/nfm/JacobsSPVPP11}
B.~Jacobs, J.~Smans, P.~Philippaerts, F.~Vogels, W.~Penninckx, and F.~Piessens.
\newblock Verifast: {A} powerful, sound, predictable, fast verifier for {C} and
  {J}ava.
\newblock In \emph{Proceedings of the Third International Symposium on {NASA}
  Formal Methods, {NFM} 2011, Pasadena, CA, USA, April 18-20, 2011.}, pages
  41--55, 2011.
\newblock \doi{10.1007/978-3-642-20398-5_4}.
\newblock URL \url{http://dx.doi.org/10.1007/978-3-642-20398-5_4}.

\bibitem[Jac(2012)]{roccc-manual}
\emph{{ROCCC 2.0 User's Manual}}.
\newblock Jacquard Computing~Inc., revision 0.74 edition, June 2012.
\newblock http://roccc.cs.ucr.edu/UserManual.pdf.

\bibitem[Klint et~al.(2009)Klint, Van Der~Storm, and Vinju]{klint2009rascal}
P.~Klint, T.~Van Der~Storm, and J.~Vinju.
\newblock Rascal: A domain specific language for source code analysis and
  manipulation.
\newblock In \emph{Source Code Analysis and Manipulation, 2009. SCAM'09. Ninth
  IEEE International Working Conference on}, pages 168--177. IEEE, 2009.

\bibitem[Lammel et~al.(2009)Lammel, Jones, and Magalhaes]{DataGenerics}
R.~Lammel, S.~P. Jones, and J.~P. Magalhaes.
\newblock {The SYB Package}.
\newblock \url{https://hackage.haskell.org/package/syb}, 2009.

\bibitem[Lindtjorn et~al.(2011)Lindtjorn, Clapp, Pell, Fu, Flynn, and
  Mencer]{lindtjorn2011-beyond-micro}
O.~Lindtjorn, R.~G. Clapp, O.~Pell, H.~Fu, M.~J. Flynn, and O.~Mencer.
\newblock {Beyond Traditional Microprocessors for Geoscience High-Performance
  Computing Applications}.
\newblock \emph{{IEEE} Micro}, 31\penalty0 (2):\penalty0 41--49, 2011.

\bibitem[{Maxeler Technologies}(2016)]{techologies16:max_compiler}
{Maxeler Technologies}.
\newblock {Max Compiler MPT}.
\newblock https://www.maxeler.com/solutions/low-latency/maxcompilermpt/, March
  2016.

\bibitem[O'Hearn et~al.(2001)O'Hearn, Reynolds, and
  Yang]{DBLP:conf/csl/OHearnRY01}
P.~W. O'Hearn, J.~C. Reynolds, and H.~Yang.
\newblock Local reasoning about programs that alter data structures.
\newblock In L.~Fribourg, editor, \emph{CSL}, volume 2142 of \emph{Lecture
  Notes in Computer Science}, pages 1--19. Springer, 2001.
\newblock ISBN 3-540-42554-3.

\bibitem[Reynolds(2002)]{DBLP:conf/lics/Reynolds02}
J.~C. Reynolds.
\newblock Separation logic: A logic for shared mutable data structures.
\newblock In \emph{LICS}, pages 55--74. IEEE Computer Society, 2002.
\newblock ISBN 0-7695-1483-9.

\bibitem[Rubio et~al.(2015)Rubio, Glass, Kuper, and de~Groote]{cluster2015}
D.~Rubio, C.~W. Glass, J.~Kuper, and R.~de~Groote.
\newblock Introducing and exploiting hierarchical structural information.
\newblock In \emph{IEEE International Conference on Cluster Computing
  (CLUSTER), 2015}, pages 777--784. IEEE, 2015.

\bibitem[Schupp et~al.(2002)Schupp, Gregor, Musser, and Liu]{Schupp2002}
S.~Schupp, D.~Gregor, D.~Musser, and S.-M. Liu.
\newblock Semantic and behavioral library transformations.
\newblock \emph{Information and Software Technology}, 44\penalty0
  (13):\penalty0 797--810, 2002.

\bibitem[van Wijngaarden and Visser(2003)]{van2003-prog-transf-mech}
J.~van Wijngaarden and E.~Visser.
\newblock {Program Transformation Mechanics}.
\newblock Technical report, Technical Report UU-CS-2003-048, Universiteit
  Utrecht, 2003.

\bibitem[Verdoolaege and Grosser(2012)]{poly_pet}
S.~Verdoolaege and T.~Grosser.
\newblock Polyhedral extraction tool.
\newblock In \emph{Second International Workshop on Polyhedral Compilation
  Techniques (IMPACT'12), Paris, France}, 2012.

\bibitem[Vigueras et~al.(2016)Vigueras, Carro, Tamarit, and
  Mari{\~n}o]{vigueras16:learning_code_trans}
G.~Vigueras, M.~Carro, S.~Tamarit, and J.~Mari{\~n}o.
\newblock {Towards Automatic Learning of Heuristics for Mechanical
  Transformations of Procedural Code}.
\newblock In \emph{First Workshop on Program Transformation for Programmability
  in Heterogeneous Architectures (PROHA'16)}, March 2016.

\bibitem[Visser(2004)]{visser04:stratego-XT-0.9}
E.~Visser.
\newblock {Program Transformation with Stratego/XT: Rules, Strategies, Tools,
  and Systems in StrategoXT-0.9}.
\newblock In C.~Lengauer, D.~Batory, C.~Consel, and M.~Odersk, editors,
  \emph{Domain-Specific Program Generation}, volume 3016 of \emph{Lecture Notes
  in Computer Science}, pages 216--238. Springer-Verlag, June 2004.

\bibitem[Visser(2005)]{Visser2005}
E.~Visser.
\newblock {A Survey of Strategies in Rule-Based Program Transformation
  Systems}.
\newblock \emph{Journal of Symbolic Computation}, 40\penalty0 (1):\penalty0
  831--873, 2005.
\newblock Special issue on Reduction Strategies in Rewriting and Programming.

\bibitem[Willadsen(2016)]{meld}
K.~Willadsen.
\newblock Meld.
\newblock http://meldmerge.org/, February 2016.
\newblock Retrieved on February 2016.

\end{thebibliography}


%
%


\end{document}